\documentclass[runningheads,a4paper]{cmmr2023}

\usepackage{times}
\usepackage{amssymb}
\usepackage[square,numbers]{natbib}
\usepackage{graphicx}
\usepackage{subcaption}
\usepackage{algorithm}
\usepackage{algpseudocode}
\usepackage{amssymb}
\usepackage{amsmath}
\usepackage{paralist}
\usepackage[hyphens]{url}
\usepackage{todonotes}
\usepackage{booktabs}
\usepackage{graphicx}
\usepackage{hyperref}
\usepackage{enumitem}
\usepackage{threeparttable}
\usepackage[english]{babel} 
\usepackage [autostyle, english = american]{csquotes}
\MakeOuterQuote{"}

\newcommand{\keywords}[1]{\par\addvspace\baselineskip
\noindent\keywordname\enspace\ignorespaces#1}

\usepackage{soul}
\usepackage{xcolor}
\usepackage{etoolbox}
\patchcmd{\thebibliography}{\list}{\vspace{-3em}\list}{}{}

\begin{document}

\mainmatter  

\title{Exploring Diverse Sounds: Identifying Outliers in a Music Corpus}

\titlerunning{Identifying Music Outliers}

%
%
\author{Le Cai\inst{1}, Sam Ferguson\inst{1}, Gengfa Fang\inst{1}, \and Hani Alshamrani\inst{1}
\thanks{We would like to acknowledge the support received from Assoc. Prof. Sam Ferguson, In Addition, Le Cai wants to thank his partner Hanyu Meng for her unwavering patience and constant presence in his life.}
}
%
\authorrunning{Cai et al.}

\institute{Creativity and Cognition Studios\\ Faculty of Engineering and IT\\ University of Technology Sydney \\ 
}

%

\maketitle

\begin{abstract}
Existing research on music recommendation systems primarily focuses on recommending similar music, thereby often neglecting diverse and distinctive musical recordings. Musical outliers can provide valuable insights due to the inherent diversity of music itself. In this paper, we explore music outliers, investigating their potential usefulness for music discovery and recommendation systems. We argue that not all outliers should be treated as irrelevant data, as they can offer unique perspectives to contribute to a richer musical understanding. We attempt to identify 'Genuine' music outliers, which may reveal unique aspects of an artist's repertoire and serve to enhance music exploration and discovery.
\keywords{Music Outlier · Music Outlier Detection · Audio characteristics · Music discovery

}
\end{abstract}

\section{Introduction}


In the field of music information retrieval, a primary focus is often on finding similarities among digital musical recordings, to enable recommendation systems and facilitate music discovery~\cite{SheikhFathollahi2021,Knees2013,Schedl2014,calcina2022measuring}. Given this context, analysis of outliers has attracted less research attention~\cite{app9030552}, as they are often considered  irrelevant data and removed during preprocessing, or are naturally scored lower by most similarity-focused algorithms~\cite{outlier2}. 
However, outliers in the context of music can provide interesting insights and reveal unique patterns, as music inherently exhibits great diversity~\cite{Panteli2017,app11052270}. 

In this paper, we explore the identification and categorization of music outliers, with an aim to ultimately enhance music discovery and recommendation systems. We propose a method to describe and discover genuine musical outliers based on audio characteristics, such as tempo and loudness. By doing so, we aim to identify outliers that can provide valuable information for music discovery while not being non-musical. We present a definition of what constitutes a 'Genuine' music outlier and investigate its characteristics. Genuine outliers exhibit unique characteristics that set them apart from an artist's main style, providing insightful information for music discovery.

This paper is structured as follows: The introduction presents the motivation of our study, followed by a comprehensive literature review, which discusses the relevant background and prior research. Next,  the aims and objectives  are followed by the methodology section which outlines our proposed definition of Genuine music outliers and the subsequent dataset and algorithm developed for the detection. The results and discussion section evaluates the effectiveness of our algorithm based on the dataset and provides insights into its performance. Finally, the conclusion summarizes our findings and highlights the implications of our work, while also suggesting potential avenues for future research in the realm of music outlier detection and analysis.
\vspace{-3mm}
\section{Related Work}

\subsection{What Makes a Song Different from Audio?}
Understanding outliers in the context of music recommendation systems necessitates a thorough examination of their diverse nature and the ability to differentiate actual music from other forms of audio. 
Mller~\cite{Mller2015} provides a comprehensive overview of music structure analysis, focusing on techniques for segmenting and organizing music into meaningful sections, laying the foundation for understanding the key aspects of music structure and demonstrating how various representations and algorithms can be used to analyze and compare music pieces.
A system for finding structural descriptions of musical pieces defines the structure of a piece as segments with specific time ranges and labels, with segments sharing the same label considered occurrences of a particular structural part~\cite{Paulus2006}. In another study, a multi-task deep learning framework is introduced for directly modeling structural semantic labels in music, such as "verses" and "choruses", from audio signals. This approach proposes a 7-class taxonomy that includes intro, verse, chorus, bridge, outro, instrumental, and silence, and consolidates annotations from four different datasets~\cite{Wang2022}. A large-scale analysis of songs in 315 different societies has found that songs share universal features like tonality, rhythm, and repetitive structures~\cite{Mehr2019}. A study conducted by Shuqi et al~\cite{Shuqi}. analyzes the significance of repetition and structure in music, specifically in popular music, and demonstrated that deep learning models often struggle to identify these essential elements, which are crucial for generating coherent and appealing musical pieces. Subsequently, Sargent et al~\cite{Sargent}. introduced a fourth principle: regularity. This principle posits that musical segments possess a certain degree of regularity, which can be leveraged to better understand and analyze the structure of the music.
\vspace{-3mm}
\subsection{Outlier Detection Approaches}
The purpose of outlier detection algorithms is to identify patterns and samples that deviate significantly from the normal characteristics of a group of data~\cite{Hawkins1980}. The reason to detect outliers is it can providing interesting insights contribute to a richer understanding of an artist’s work. General outlier detection methods can be categorized into 4 categories based on an overview conducted by~\cite{smiticritical2020}:
\begin{description}
\item[\textbf{Clustering-based methods:} ]use a clustering algorithm to classify the majority of the elements of the set, while also clearly defining the outlying elements of the set.  
\item[\textbf{Density-based methods:} ] identifying outliers as points in low-density regions within a data set.
\item[\textbf{Distance-based methods:}] determining the distance between points, and considering outliers to be points with a large distance from their nearest neighbors.
\item[\textbf{Statistical methods:}] using measures such as mean, median and standard deviation to identify data points that fall outside of a defined range.
\end{description}

Clustering-based methods, exemplified by~\cite{Ariyaratne2012}, utilize subspace cluster analysis to construct classification trees while addressing dataset scarcity. A deep learning model using a Clustering Augmented Learning Method (CALM) classifier improves genre classification by extracting deep time series features~\cite{Ghosal2020NovelAT}. In contrast, density-based methods like DBSCAN~\cite{DBSCAN} and OPTICS~\cite{Ankerst1999} identify outliers in low-density regions within a dataset, as demonstrated by the OPTICS algorithm applied to traditional Chinese folk music~\cite{Zhang2021}. In the exploration of automatic outlier detection methods on music genre datasets, Lu et al~\cite{Lu2016AutomaticOD} characterized outliers using their musical attributes, demonstrating the potential of these techniques to unveil unique insights into music structure and diversity within genre classification. Distance-based methods, including K-means~\cite{Azcarraga2017} and CLARANS~\cite{1033770}, examine the distance between data points and their nearest neighbors, considering outliers as points with large distances from the nearest neighbors. This approach has been successfully applied to traditional Irish music in~\cite{Shingte2019UnsupervisedLA}. Meanwhile, statistical-based methods employ measures such as mean, median, and standard deviation to identify data points outside a defined range to reveal patterns in cluster structure dynamics in popular music data~\cite{Singh2022}. However, while these studies focus on detecting outliers based on their statistical properties, the potential of the outliers themselves for music discovery has not been as extensively investigated.


\vspace{-3mm}
\section{Aims \& Objectives}
In this paper, our aims center around exploring the potential  of music outliers for music discovery. To achieve this, we propose the following objectives:
\begin{description}
    \item \textbf{Propose an approach to describe musical outliers:} by examining various attributes that distinguish them from an artist's typical style, facilitating their identification and analysis. This implies establishing a clear definition to describe 'Genuine' music outliers as a distinct category of outliers that exhibit meaningful deviations from a set of existing digital musical recordings.
    \item \textbf{Categorise music outliers:} into meaningful categories based on their distinguishing characteristics to create a meaningful interpretation, that also can help find the outliers that are helpful to understanding and discovering interesting music while excluding outliers that hold little data. 
\end{description}
\vspace{-3mm}
\section{Method}

We introduce the concept of genuine outliers within the context of music data and explore their potential value in recommendation systems. To accomplish this, we first propose a definition for genuine outliers, then create a labelled dataset for evaluation, and finally, apply an outlier detection algorithm to validate our definition.

\subsection{Definition of Genuine Music Outliers}

\begin{figure*}[t!]
        \subfloat[Example of a subsequence of audio forms a complete song]{%
            \includegraphics[width=.4\linewidth]{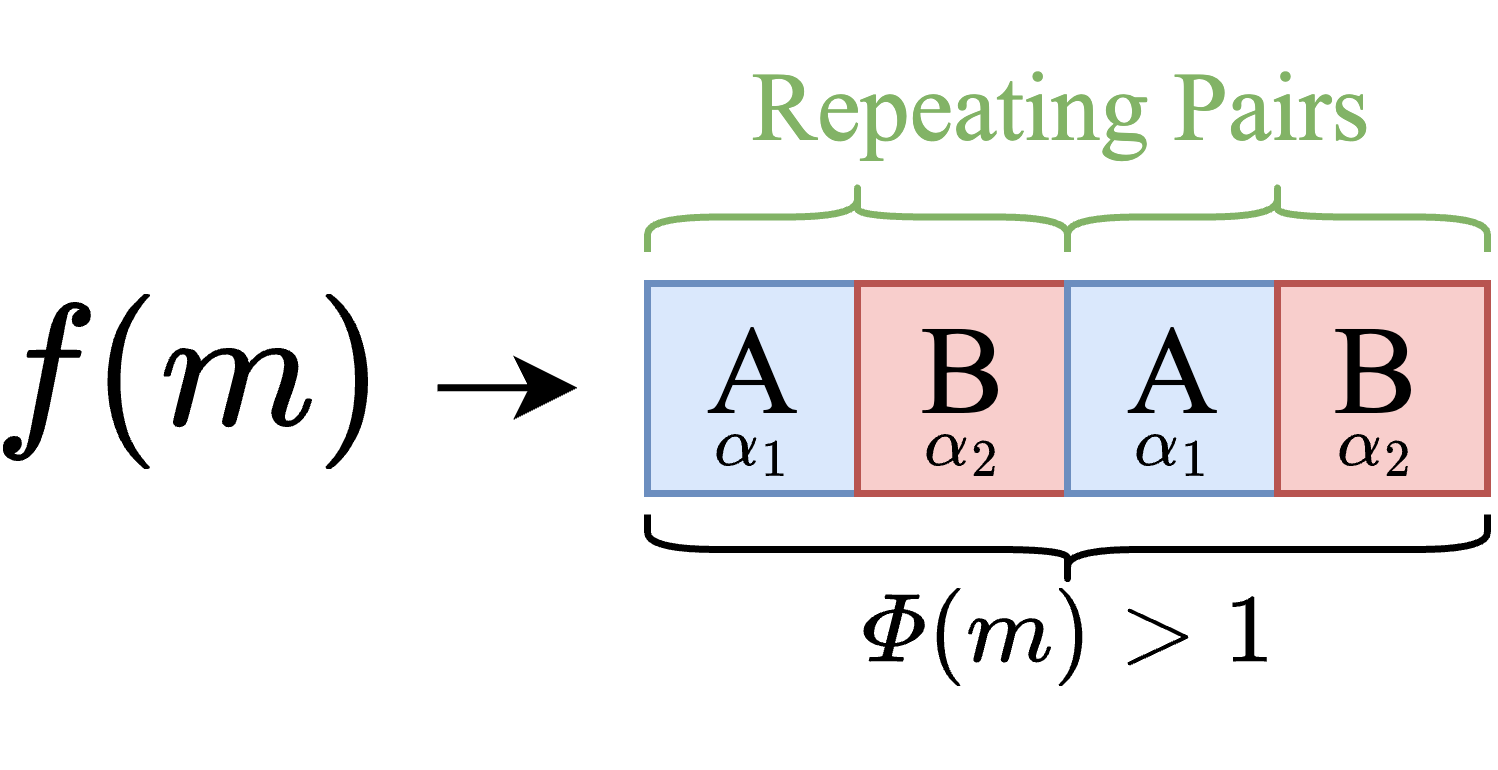}%
            \label{subfig:complete_form}%
        }\hfill
        \subfloat[Example of a k-means cluster with identified outlier]{%
            \includegraphics[width=.4\linewidth]{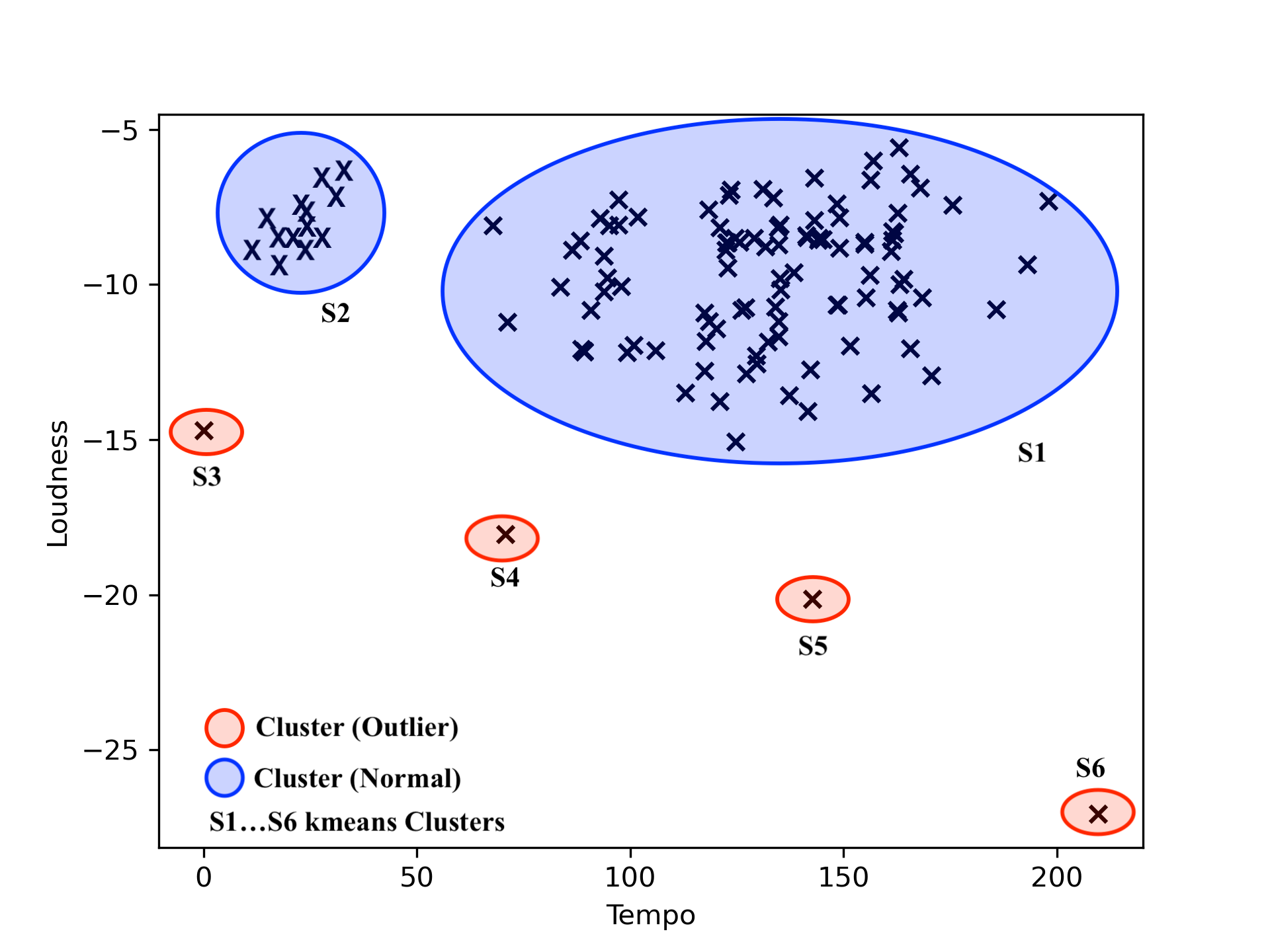}%
            \label{subfig:distinctivenetss}%
        }
        \caption{The Definition of Genuine Music Outliers}
        \label{fig:artist_kde}
\end{figure*}

A "Genuine" music outlier is a complete song that maintains an artist's typical musical structure while distinctly diverging in sound and style from their predominant body of work, due to the complex nature of music, in our scenario, we consider pop music only.

To achieve this, a genuine outlier must satisfy the following constraints: 1): Forms a complete song, this distinguishes the identified outlier must be a song, not something else, e.g. speech, or sound effect, shown in fig.\ref{subfig:complete_form}. To ensure this, audio must satisfy these conditions: a): The length of music structure $\phi$ must be greater than one in a subsequence, this means the input recordings must at least have one or more music structures otherwise it is not music.
b): The identified music structure must at least has one unique pair. e.g. a music structure in $(A-B)$ is a song that can identify at least one unique pair, not $(A-A)$ or $(B-B)$. c): The identified music structure must contain repeated parts, as repetition is a crucial element in music. A piece of audio that forms music should exhibit repeating sections, such as $(A-B-A-B)$ patterns. The formal definition of a Genuine outlier is as follows:

Let $M$ represents the set of all recordings produced by a specific artist. For a finite set $S_{M}$, suppose function $\Phi: M\rightarrow \mathbb{N^{+}}$ that maps a recording to a finite integer and a function $f_{\Phi}: M\rightarrow S^{\Phi(M)}_{M}$ that maps a recording $m\in M$ to a finite length sequence.
A recording $m \in M$ is defined as a "Genuine" music outlier if the following conditions is satisfied:

\textbf{1. Forms A Complete Song:} 
For any recording $m \in M$, the sequence $f_{\Phi}(m)$ and number $\Phi(m)$ should satisfy the following:
\begin{enumerate}
    \item $\Phi(m) > 1$,
    \item There exist a subsequence $a_{k_{1}}...a_{k_{p}}(k_{1}<...<k_{p},p>1)$ from $f_{\Phi}(m)$ that, at least one repeated subsequence from $f(m)$ can be found, i.e., there exist $a_{k^{'}_{1}}...a_{k^{'}_{p}}(k_{1}^{'}<...<k_{p}^{'},p>1)$ that $k_{1} \neq k_{1}^{'}$ and $k_{p} \neq k_{p}^{'}$.
\end{enumerate}

For any given recording $m\in M$, if the previous condition satisfied, for given constant $C_{G}\in (0,1]$, $\kappa$ and a positive integer $N_{d}$, either of the following conditions must be satisfied,

\textbf{2. Distinctiveness:} Define $F: M \to \mathbb{R}^n$ be a function that maps each recording to an $n$-dimensional feature space, where $n$ is a positive integer representing the number of musical features being considered. Define $\text{Card}(\cdot)$ as the Cardinality of a set, $\|\cdot\|$ as the 2-norm and $\kappa$-means cluster sets $\boldsymbol{\Omega} = \{\Omega_{1},...,\Omega_{\kappa}\}(1\leq \kappa \leq \text{Card}(M))$ that

\begin{equation}
\centering
\begin{split}
    &\qquad \boldsymbol{\Omega}\in \arg\min_{\boldsymbol{\bar{\Omega}}} \sum_{i=1}^{\kappa} \sum_{\mathbf{x}\in \bar{\Omega}_{i}\subset M} \|F(\mathbf{x}) - \boldsymbol{\mu}_{i}\|^2\\
    &\text{s.t.} \quad \boldsymbol{\mu}_{i} = \frac{1}{\text{Card}(\bar{\Omega}_{i})}\sum_{\mathbf{y}\in \bar{\Omega}_{i}\subset M} F(\mathbf{y}), \text{for $1\leq i\leq \kappa$}.
\end{split}
\end{equation}

For any given integer $N_{d} > 0$, for any $m\in M$, if $m\in \Omega_{i}$ and $\text{Card}(\Omega_{i}) < N_{d}$, we say that recording $m$ is distinct with respect to $M$.


\textbf{3. Non-adherence:} 
For a given positive number $C_{G} \in (0,1]$. We define 
\begin{equation}
    R_{M}(m) := \frac{\text{Card}(\{m^{'} | f_{\Phi}(m^{'}) = f_{\Phi}(m) \text{ for all $m^{'} \in M$}\})}{\text{Card}(M)}.
\end{equation}
If $R_{M}(m) < C_{G}$, we say $m$ is not adhere to set $M$.

In conclusion, if recording $m$ does form a complete song, either the song $m$ is distinct with respect to $M$ or $m$ is not adhere to set $M$, with given $C_{G}$, $\kappa$, and $N_{d}$, then we say song $m$ is an outlier.

\subsection{Dataset Creation and Outlier Selection Algorithm}

\begin{figure}[t]
\centering
\includegraphics[width=0.9\linewidth]{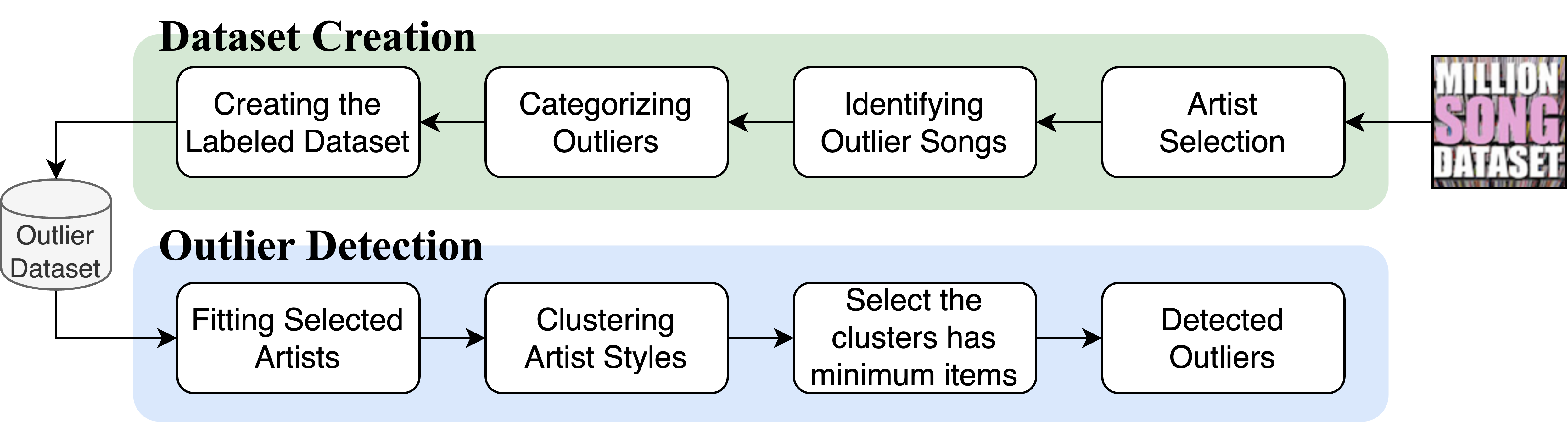}
\caption{Dataset Creation Process}
\label{fig:datacreate}
\end{figure}

In order to evaluate our above-proposed definition for Genuine music outliers, we created a dataset specifically designed to examine our hypothesis, To achieve this goal, the following steps were carried out:

 \textbf{Artist Selection:} From the Million Song Dataset \footnote{Million Song Dataset~\cite{BertinMahieux2011TheMS} is available at: \url{http://millionsongdataset.com}}, a smaller sample consisting of 10,000 songs extracted from the MSD, we randomly chose 20 artists using a random number generator to pick artist IDs. For each of the selected artists, we identified a list of all the songs contained in the dataset with the particular artist id.

\textbf{Identifying Outlier Songs:} To identify and categorize outliers within the selected artists' music, we conducted a manual listening and labeling process for these songs by ear. First, we acquainted the artist's typical styles. To achieve that, we listened to about 20\%-30\% songs in this artist's main cluster.
    After establishing the primary styles, we listened to these songs once with selected artists again, focusing on identifying tracks that significantly deviated from the typical style. Attention was paid to musical elements, such as tempo, melody, harmony, instrumentation, and song structure.
    
\textbf{Categorizing Outliers:} We then listened to the characteristics of the identified outliers, focusing on specific attributes such as tempo, melody, harmony, instrumentation, and song structure that differentiate them from the artist's dominant style. By examining these properties, we classified outliers into five categories: Error, Speech, Intro, Sound Effect, and Genuine. Each identified outlier was assigned to one of the categories based on a two-step process. First, we selected songs with a Euclidean distance greater than 3 times the z-score threshold from the main cluster. Second, we manually reviewed these outliers, focusing on listening to their distinctive features such as tempo, loudness, melody, and the composition of instruments.

\textbf{Creating the Labeled Dataset:} Finally, After identifying and categorizing outliers, we compiled a list of all songs from the selected artists, along with their corresponding outlier categories. This list included each song's title, artist, genre, and other metadata. Utilizing the pre-extracted audio features from the MSD Subset, such as tempo, loudness, key, and mode, we created a dataset compiled with data including outlier categories and the MSD pre-extracted features.






We consider this to be a classification problem where each artist typically consists of 1 main distinct style. Note we only considered the case for 1. Distinctiveness and 2. Non-adherences for the current approach, the case of forms a complete song is discarded due to the complexity of the anlysis of music structures.

Let $M$ be a set of recordings under an artist. Let $x_{i}\in \mathbb{R}^{d}$ to denote a $d$-dimensional feature vector for the $i$-th song from $M$($1\leq i\leq \mathrm{Card}(M)$). Suppose $K$ is an integer that $1 \leq K \leq \mathrm{Card}(M)$. For $1 \leq k,k^{'} \leq K$, define $C_{k}$ as a subset of $M$, such that $C_{k} \cap C_{k^{'}} = \emptyset$ for $k\neq k^{'}$, and $\cup_{1 \leq i \leq K} C_{i} = M$. In our case, $d = 2$. We adopt k-means algorithm~\cite{kmeans} to partition set $M$ that satisfies such conditions. In the next section, we first delineate the outlier categorization, followed by the presentation of outlier detection results.

\section{Results}
\subsection{Outlier Categorisations}

\begin{figure}[t]
\centering
\includegraphics[trim={1cm 1cm 1cm 1cm},clip,width=0.7\linewidth]{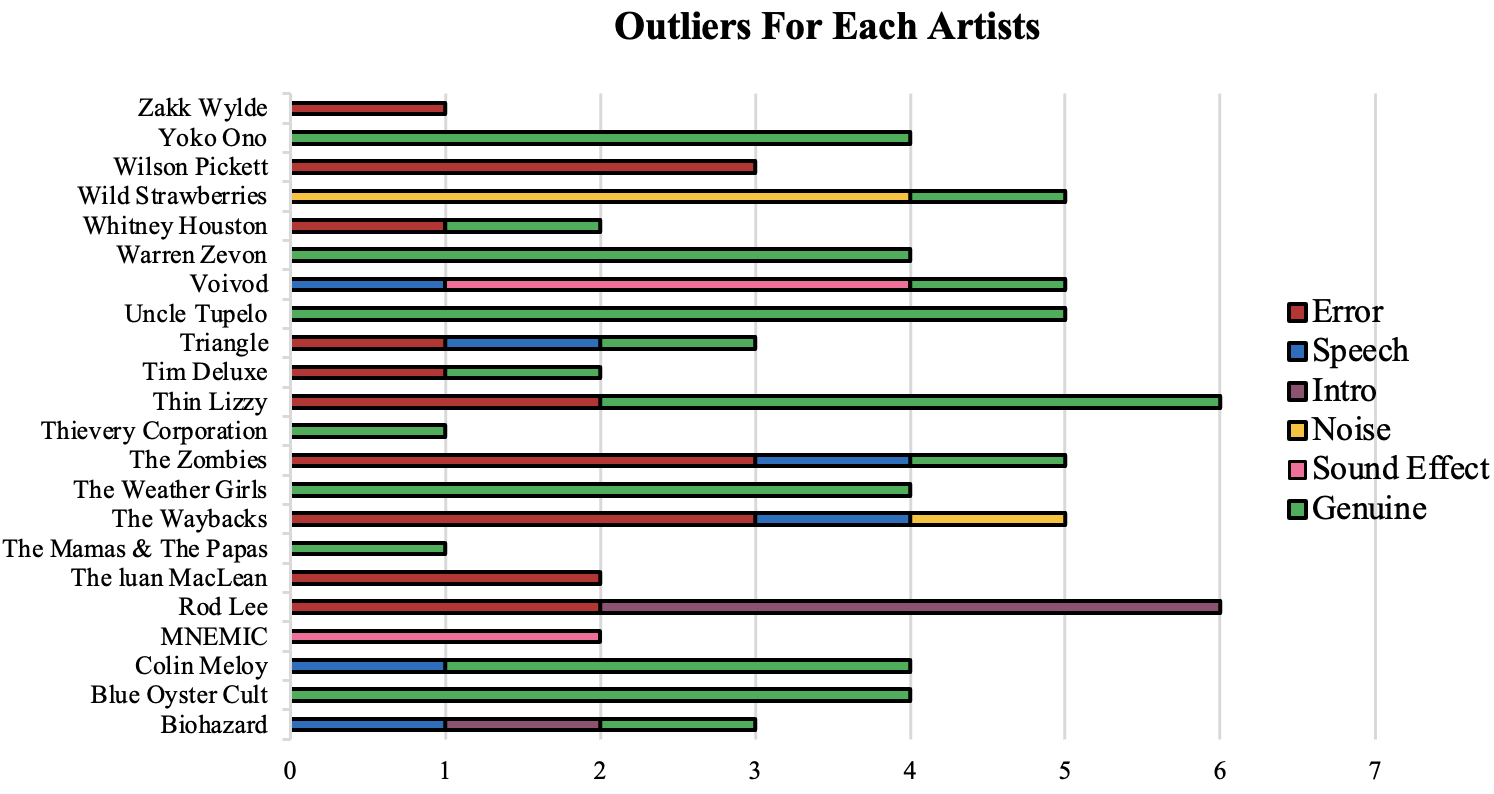}
\caption{Outlier Dataset: Categorisation of Outliers by Artists}
\label{fig:barchartofoutliers}
\end{figure}
We found 34 genuine outliers from 29 artists in 320 songs as well as 17 errors, 5 speeches, 1 intro, 5 noise, and 3 sound effects from 29 artists shown in Fig. \ref{fig:barchartofoutliers}, they can categorize into 5 types:
\begin{inparaenum}[\itshape a)\upshape]
    \item \textbf{Error:} such as erroneous data value, e.g. doubling tempo due to feature extractor failure, incorrect data type, such as text entered as a numeric type.
    \item \textbf{Noise:} unintentional sounds such as non-musical noise, jitter, or glitches in a recording, often due to recording issues or equipment quality. e.g. a live performance recording with excessive audience noise.
    \item \textbf{Speech:} such as some spoken words, or a short story integrated into the music to enhance the atmosphere of the listening experience. these types of practices commonly occur in certain music genres, such as hip and electronic music. 
    \item \textbf{Sound Effect:} This type of recording is used to create a certain mood, it can build tension and add extra impact to certain sections of the song. Particularly, sound effect is used in certain genres of music such as electronic \& rock music to add extra depth and interest to music.
    \item \textbf{Intro:}  A very short track is often less than 30 seconds in length and serves as an introduction or a brief to establish the identity of the album. 
    \item \textbf{Genuine:}  a musical piece that deviates significantly from an artist's typical style or the norm within a genre, exhibiting unique characteristics. These outliers are not classified as errors, noise, or other non-musical categories.
\end{inparaenum}
\subsection{Automation Outlier Detection Result}
\begin{table}[t]
    \centering
    \caption{The Result of Automatic Outlier Detection}
    \label{fig:table_MetricoutlierDetection}
    \small
    \setlength{\tabcolsep}{4pt} 
    \renewcommand{\arraystretch}{0.3} 
    \begin{tabular}{lllll}
    \toprule
    Artist Name            & TPR   & FPR   & TNR   & FNR   \\
    \midrule
    Zakk Wylde             & 0     & 0.455 & 0.545 & 1     \\
    Blue Oyster Cult       & 1     & 0     & 1     & 0     \\
    Biohazard              & 0.5   & 0.381 & 0.619 & 0.5   \\
    Yoko Ono               & 0.25  & 0.049 & 0.951 & 0.75  \\
    Wilson Pickett         & 1     & 0.316 & 0.684 & 0     \\
    Wild Strawberries      & 1     & 0.087 & 0.913 & 0     \\
    Whitney Houston        & 0.5   & 0     & 1     & 0.5   \\
    Warren Zevon           & 1     & 0.356 & 0.644 & 0     \\
    Voivod                 & 0.833 & 0     & 1     & 0.167 \\
    Uncle Tupelo           & 0.4   & 0.217 & 0.783 & 0.6   \\
    Triangle               & 0.75  & 0     & 1     & 0.25  \\
    Tim Deluxe             & 1     & 1     & 0     & 0     \\
    Thin Lizzy             & 0.429 & 0.37  & 0.63  & 0.571 \\
    Thievery Corporation   & 0.667 & 0.299 & 0.701 & 0.333 \\
    The Zombies            & 1     & 0.378 & 0.622 & 0     \\
    The Weather Girls      & 1     & 0.344 & 0.656 & 0     \\
    The Waybacks           & 0.5   & 0.214 & 0.786 & 0.5   \\
    The Trammps            & N/A   & 0.136 & 0.864 & N/A   \\
    The Subhumans          & N/A   & 0.133 & 0.867 & N/A   \\
    The Skatalites         & N/A   & 0.317 & 0.683 & N/A   \\
    THERION                & 0     & 0.275 & 0.725 & 1     \\
    The Mutton Birds       & N/A   & 0.269 & 0.731 & N/A   \\
    The Mission            & N/A   & 0.324 & 0.676 & N/A   \\
    The Mamas \& The Papas & 1     & 0.206 & 0.794 & 0     \\
    The Juan MacLean       & N/A   & 0.324 & 0.676 & N/A   \\
    Zee Avi                & N/A   & 0.231 & 0.769 & N/A   \\
    MNEMIC                 & 1     & 0.043 & 0.957 & 0     \\
    Rod Lee                & 1     & 0.167 & 0.833 & 0     \\
    Colin Meloy            & 0.333 & 0.062 & 0.938 & 0.667 \\
    \bottomrule
    \end{tabular}
\end{table}
The outlier detection algorithm was applied to the dataset, and the results obtained were analyzed to assess the performance of the method. The table \ref{fig:table_MetricoutlierDetection} summarizes the performance of the outlier detection for various artists in the dataset. \textbf{True Positive Rate (TPR)} and \textbf{False Positive Rate (FPR)} represent the proportion of outliers that were correctly and incorrectly identified as outliers, respectively.  \textbf{True Negative Rate (TNR)} and \textbf{False Negative Rate (FNR)} represent the proportion of non-outliers  that were correctly and incorrectly identified as non-outliers, respectively.  \textbf{Not Applicable (N/A):} indicates no outliers being identified for a particular artist.

\section{Discussion}
\subsection{Outlier Categorisation}
\begin{figure}[t]
\centering
\includegraphics[width=0.5\linewidth]{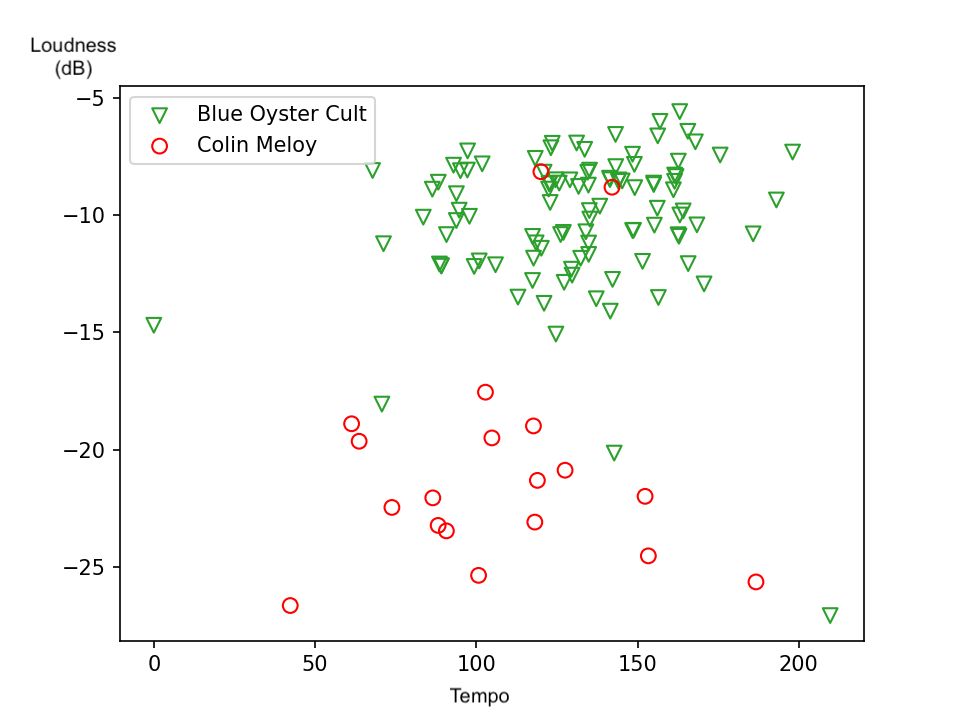}
\caption{Outlier Overlap: Shared characteristics between two artists' outliers.}
\label{fig:makeaclusteryfigure}
\end{figure}
Genuine outliers are complete songs that adhere to an artist's typical musical structure but differ in sound and style. Often custom-made for specific events, they incorporate unique elements like distinct percussion. For instance, Colin Meloy's "Lazy Little Ada" is situated at the center of Blue Oyster Cult's cluster (Figure \ref{fig:makeaclusteryfigure}), deviating from Meloy's usual style yet resembling Blue Oyster Cult's. In comparison, Meloy's cluster exhibits lower loudness, potentially due to less percussion. When percussion is added to Meloy's outliers, they show similarities to Blue Oyster Cult's cluster. Conversely, most of Blue Oyster Cult's songs form a well-defined cluster, but 4 outliers incorporate synthesizers and lack percussion,  producing similarity to Meloy's cluster.

In contrast, non-genuine outliers can be categorized into four types: Error, Speech, Sound Effect, and Intro. 1) Error tracks result from feature extractor misinterpretations or exceptions, such as doubling tempo or missing audio features. 2) Speech is often used in albums for storytelling, setting narrative themes, or as interludes to create a narrative flow between musical pieces, e.g., Kendrick Lamar's "To Pimp a Butterfly." 3) Sound effects enhance the musical narrative, especially in concept albums like Pink Floyd's "The Trial." 4) Intros are short tracks that set the tone, introduce themes, or provide transitions between songs. However, these tracks do not follow typical musical structures (e.g., ABAB, ABAC), so we consider them non-genuine outliers.

\vspace{-2mm}
\subsection{Outlier Detection Result}
\vspace{-0mm}
The overall results of the outlier detection algorithm show that it is capable of identifying outliers in an artist's body of work. The algorithm performs exceptionally well on some artists by correctly classifying all outliers and non-outliers. However, varying performance among artists suggests the need for refining the algorithm with additional constraints. In summary, we can obtain these insights from the following results: 

The consideration of distinctiveness in the algorithm has worked well on artists that primarily have one style, for example, Blue Oyster Cult, The Mamas \& The Papas, and Wild Strawberries had a perfect TPR and TNR, indicating that the definition works well for this artist even without considering the constraints of whether it forms a complete song. Furthermore, we find the distinctiveness constraint effective in identifying genuine music outliers in some artists. For example, the algorithm isolated outliers in artists such as Blue Oyster Cult (TPR: 1.0, FPR: 0, TNR: 1, FNR: 0) and Warren Zevon (TPR: 1.0, FPR: 0.356, TNR: 0.644, FNR: 0).

However, the varied performances across different artists suggest that discarding the consideration of constraints 1. Forms A Complete Song and 2. Non-adherence may lead to the algorithm's inability to accurately determine whether an outlier is genuine, as it lacks the capacity to assess the music structure of input recordings. The reasons causing these varying performances can be summarized as follows:

 \textbf{Mixed content:} Recordings may contain various non-musical elements, including intros with predominantly speech content (e.g., "Rod Intro" by Rod Lee), live recordings featuring audience applause or speech interactions with the audience (e.g., "Dracula's Daughter" by Colin Meloy), or studio chats consisting solely of speech (e.g., "The Way I Feel Inside / Studio Chat" by The Zombies). These non-musical segments may introduce noise and impact the algorithm's performance.

 \textbf{Noise and artifacts:} The presence of noise, artifacts, or other non-musical elements within a recording might lead to it being classified as an outlier, even if it does not constitute a genuine outlier in terms of musical content. Such factors can interfere with the algorithm's capability to accurately assess a song's structure.

 \textbf{Transitional pieces:} Some artists release tracks with transition pieces containing sound effects or ambient sounds, which can be difficult for the algorithm to categorize as genuine outliers. Notable examples include "Catalepsy I" by Voivod and "The Audio Injection" by MNEMIC.

\begin{figure}[t]
        \subfloat[Artist (Blue Oyster Cult)]{%
        \includegraphics[width=0.45\linewidth]{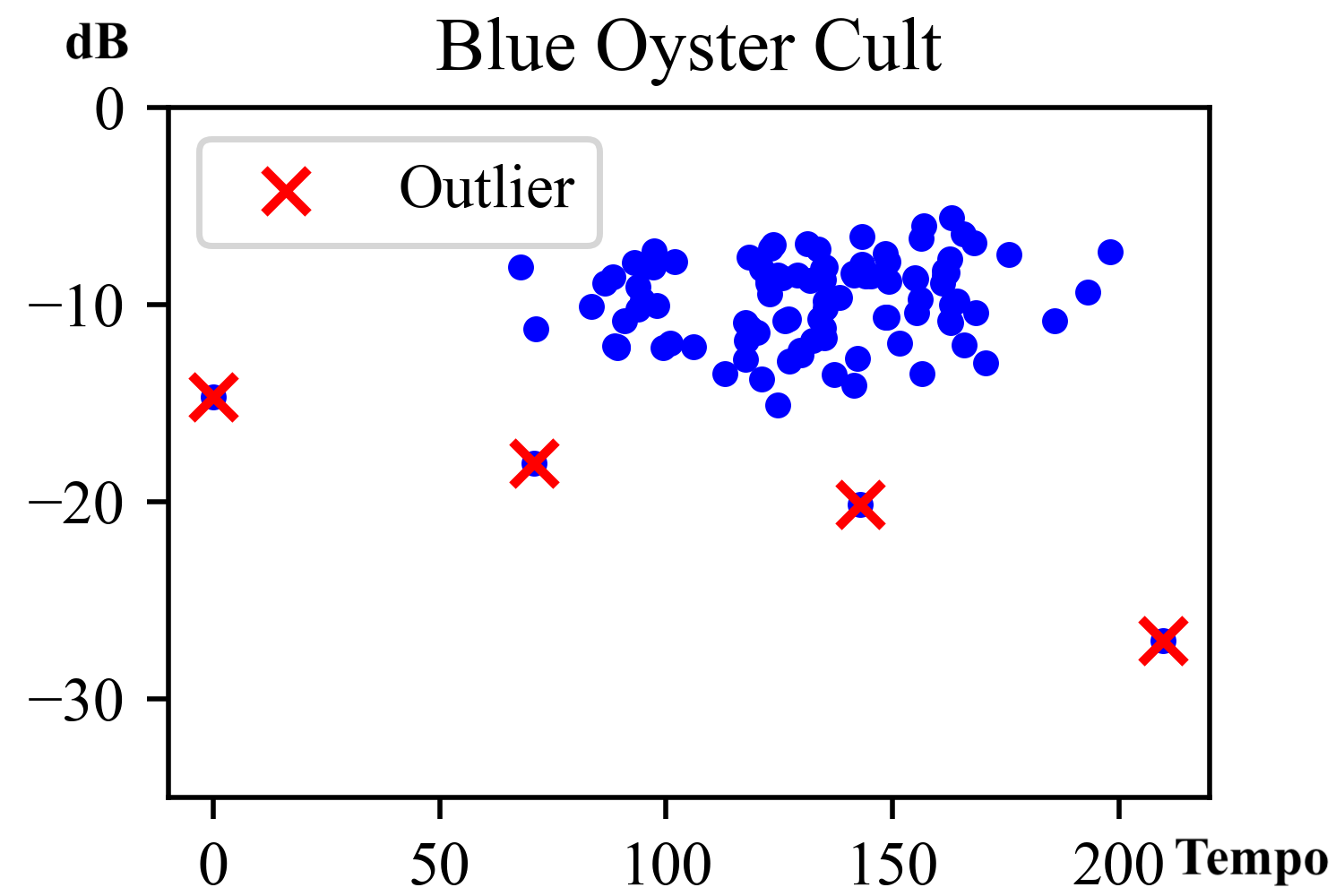}%
            \label{subfig:a_1}%
        }\hfill
        \subfloat[Artist (Rod Lee)]{%
            \includegraphics[width=0.45\linewidth]{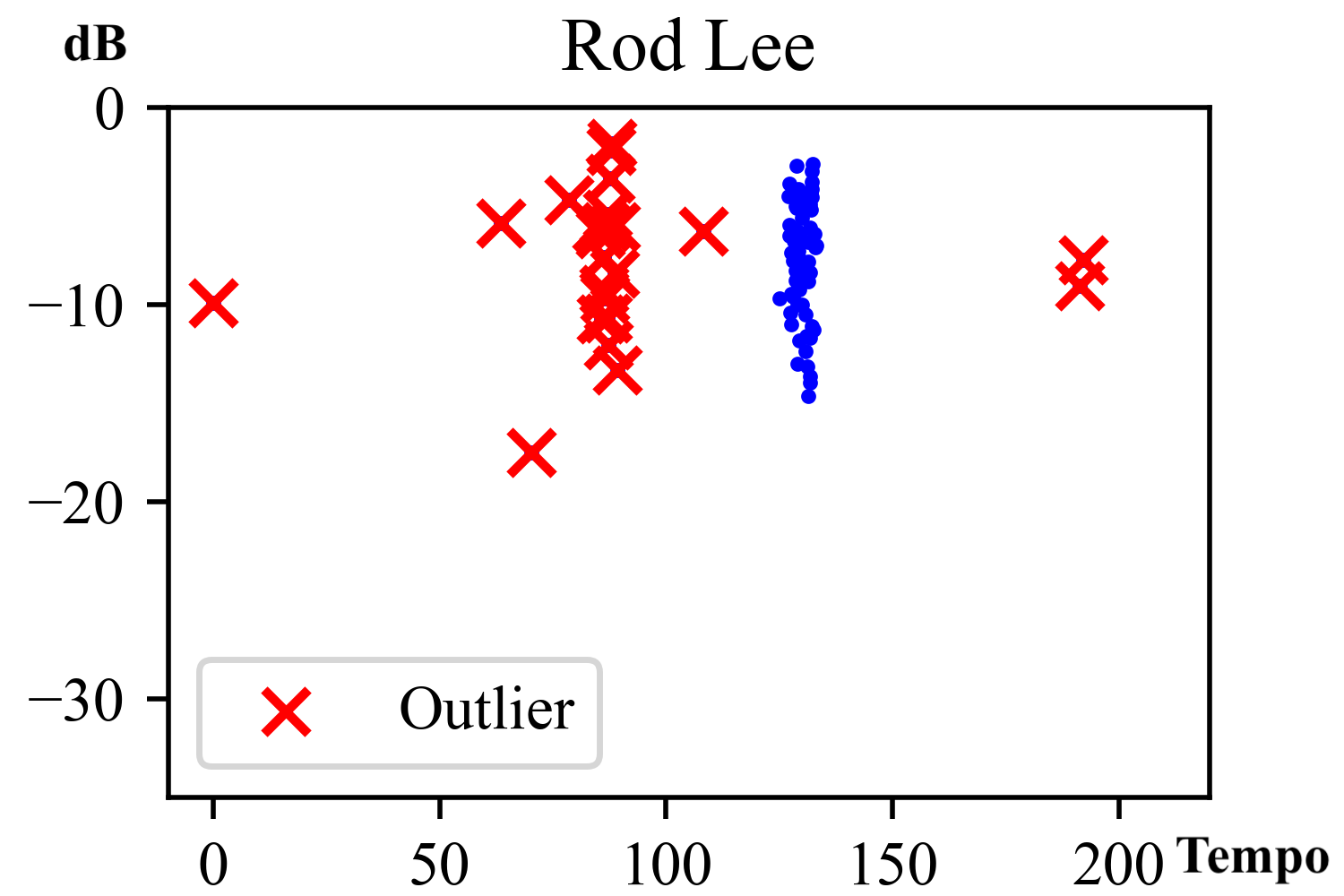}%
            \label{subfig:b_1}%
        }\\
        \caption{Automatic Outlier Detection Using $k$-means: Demonstrating the algorithm works well on single-style artists (a) but fails for multi-style artists (b).}
        \label{fig:outlierDetectionResult}

\end{figure}

\vspace{-3mm}

\subsection{Limitations and Future Work}
\vspace{-2mm}

Overall, the findings of this study highlight the importance of considering outliers and their potential impact on the audio characteristics of different genres and artists in music production. It suggests several avenues for future research, including: 

\textbf{Integration of more music features:} Expanding the range of music features integrated into the algorithm can potentially improve its performance in detecting genuine outliers. The current features, such as loudness and tempo, may not fully capture the characteristics and the style of music. Therefore, incorporating additional features like timbre, harmony, and chroma could enhance the algorithm's effectiveness in distinguishing genuine outliers from the rest of an artist's work.
    
\textbf{Consideration of "Forms A Complete Song" and "Non-Adherence" constraints:} Incorporating the "Forms A Complete Song" constraint ensures that the detected outliers are actual pieces of music. This guarantees that the outliers are indeed genuine musical outliers and not artifacts or other irrelevant audio content. The 'Non-Adherence' constraint ensures detected outliers distinctly deviate from an artist's typical musical structure. This helps to identify unique songs that stand out from an artist's typical style. Furthermore, music segmentation techniques can be considered to extract musical parts from audio pieces that may contain both speech and music. This will aid in ensuring the detected item actually is music.

\textbf{Handling artists with more than one style:} The current outlier detection approach may struggle with artists exhibiting multiple styles, such as Rod Lee (shown in fig.\ref{fig:outlierDetectionResult}). Addressing this limitation would result in a more representative understanding of an artist's work and improve the accuracy of outlier detection. One possible solution is to consider an artist's stylistic diversity when detecting outliers, thereby accounting for the various styles present within their body of work.
     
\textbf{Exploration of other clustering models:} In this study, only the k-means clustering algorithm was considered for the clustering model. However, k-means is based on circular data, which can lead to suboptimal results. Exploring other clustering models that may better handle the complexities of music data could further enhance the performance of the outlier detection algorithm and  yield more accurate results.

\section{Conclusion}

In conclusion, this study has proposed a definition for genuine music outliers and explored the application of an outlier detection algorithm in music genre datasets. The results have demonstrated that the consideration of distinctiveness is a reasonable starting point for detecting music outliers. However, the current approach lacks the ability to detect the music structure and struggles when handling artists with more than one style. To overcome these limitations, future work should focus on integrating more music features, such as timbre, harmony, and chroma, and considering the constraints of "Forms A Complete Song" and "Non-Adherence." Furthermore, music segmentation techniques should be explored to extract musical parts from audio pieces containing both speech and music. Handling artists with multiple styles and exploring alternative clustering models, such as those that can better accommodate non-circular data, are other avenues for improvement. By addressing these limitations and incorporating these suggestions, the proposed outlier detection approach can be further refined and made more robust for detecting genuine music outliers in diverse music genre datasets.
\vspace{-2mm}

\makeatletter
\renewenvironment{thebibliography}[1]
     {\section*{\refname}%
      \@mkboth{\MakeUppercase\refname}{\MakeUppercase\refname}%
      \list{\@biblabel{\@arabic\c@enumiv}}%
           {\settowidth\labelwidth{\@biblabel{#1}}%
            \leftmargin\labelwidth
            \advance\leftmargin\labelsep
            \@openbib@code
            \usecounter{enumiv}%
            \let\p@enumiv\@empty
            \renewcommand\theenumiv{\@arabic\c@enumiv}}%
      \sloppy
      \clubpenalty4000
      \@clubpenalty \clubpenalty
      \widowpenalty4000%
      \sfcode`\.\@m}
     {\def\@noitemerr
       {\@latex@warning{Empty `thebibliography' environment}}%
      \endlist}
\makeatother
\bibliographystyle{splncs04}
\footnotesize \bibliography{cmmr2023_template}
\end{document}